# Analysis of Null Related Beampattern Measures and Signal Quantization Effects for Linear Differential Microphone Arrays

Shweta Pal, Arun Kumar, Monika Agrawal

*Abstract*— A differential microphone array (DMA) offers enhanced capabilities to obtain sharp nulls at the cost of relatively broad peaks in the beam power pattern. This can be used for applications that require nullification or attenuation of interfering sources. To the best of our knowledge, the existing literature lacks measures that directly assess the efficacy of nulls, and null-related measures have not been investigated in the context of differential microphone arrays (DMAs). This paper offers new insights about the utility of DMAs by proposing measures that characterize the nulls in their beam power patterns. We investigate the performance of differential beamformers by presenting and evaluating null-related measures namely null depth (ND) and Null Width (NW) as a function of depth level relative to the beam power pattern maxima. A study of signal quantization effects due to data acquisition for 1$^{st}$, 2$^{nd}$ and 3$^{rd}$ order linear DMAs and for different beampatterns i.e. dipole, cardioid, hypercardioid and supercardioid is presented. An analytical expression for the quantized beamformed output for any general $N^{th}$ order DMA is formulated. Simulation results of the variation of ND with number of quantization bits and the variation of NW as a function of depth are also presented and inferences are drawn. Lab experiments are conducted in a fully anechoic room to support the simulation results. The measured beampattern exhibits a pronounced null depth, confirming the effectiveness of the experimental setup.

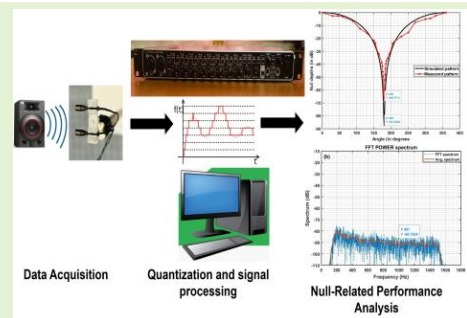

## I. Introduction

Speech and audio transmissions often suffer significant degradation from noise, interference, and reverberation.

Using a single microphone limits spatial realism, as it cannot capture directional sound, and restricts post-processing flexibility, making it difficult to effectively separate the desired signal from noise, reverberation, or competing sources. This confines signal enhancement techniques to a narrow set of options, reducing overall audio quality and control. Microphone arrays have been introduced to address these limitations, enabling improved signal acquisition and enhanced processing capabilities in both single- and multi-source noisy environments. [1], [2]. A microphone array is a structured arrangement of multiple omnidirectional or directional sensors that collectively sample the sound field with spatial diversity [3]. Differential Microphone Arrays (DMAs) are special arrays that combine closely spaced sensor outputs in differential/subtractive manner such that the microphone array difference output is proportional to the spatial derivative of an acoustic pressure field [3], [4]. DMAs are designed to have a compact size due to the intrinsic assumption in DMA that microphones must be positioned close enough for the true acoustic pressure differentials to be accurately approximated by the finite differences between their outputs. Accurate estimation requires that the inter-element spacing be significantly smaller than the acoustic wavelength across the target frequency range. This small-spacing constraint provides both the primary strength and limitation of differential arrays as they offer high directional gain with small size, frequency-invariant beampatterns, but at the cost of high white noise amplification, making them highly susceptible to self-noise and errors from microphone amplitude and phase mismatches [5]-[8].

Presently, DMAs have been mostly examined using traditional performance measures such as directivity factor (DF), which is relevant for the accuracy of finding target source directions through the peaks in beam power pattern plots [9]-[10]. (We will henceforth refer to the beam power pattern as beampattern in the paper). However, in differential processing, the more important aspect is the placement of nulls, which are relatively sharp rather than peaks, that tend to be spatially broad. Applications that require effective suppression of interfering sources from specific directions by placing sharp nulls in those directions can use DMAs that are particularly well-suited. Thus, measures that characterize these nulls of beampattern are of primary importance. Examples include directional noise suppression, such as engine noise suppression in car infotainment systems and interfering source suppression in differential hearing aids.

The existing measures for DMAs given in literature, such as

Shweta Pal, Arun Kumar, and Monika Agrawal are with Centre for Applied Research in Electronics, IIT Delhi-110016, India (email: Shweta.Pal@care.iitd.ac.in, arunkm@care.iitd.ac.in) and maggarwal@care.iitd.ac.in
(Corresponding author: Shweta Pal)

DF, white noise gain (WNG), Front-to-back-ratio (FBR) [3] etc. are not primarily indicative of the efficacy of the beampattern nulls. These measures are about the broad beampattern peaks rather than about characterizing the nulls. Authors have studied the directivity of a $1^{st}$ order steerable square differential array and about the directivity patterns of a linear microphone array by exploiting design aspects of a DMA [11]-[12]. Authors in [13], applied MVDR beamforming algorithm to a linear DMA geometry for DOA estimation of both narrowband and broadband sources by detecting peaks in the beampattern of the array. In one study, Wang et al. [14] investigated how changing null directions might affect beamforming performance, particularly DF, signal-to-interference noise ratio, etc. In [15], the authors present a new design method for non-uniform linear DMAs that effectively eliminates high-frequency nulls in the mainlobe while ensuring frequency invariance for DF and mainlobe width. The authors in [16] discussed nulls in the context of a high-speed antenna platform by presenting a novel null broadening beamforming strategy to mitigate the degradation of interference suppression resulting from rapid jammer movement. In an initial work [17], quantization analysis for a two-microphone DMA geometry has been explored by examining a number of performance measures, including beampattern, DF, front-to-back ratio (FBR), etc.

To the best of our knowledge, null-related measures, which are important for applications that aim to explicitly attenuate interfering sources, have not been explored in the context of DMAs. This paper recasts the utility of DMAs in the context of suppression of interference sources, alternatively on the importance of nulls in its beampattern. This paper introduces and defines null-related measures such as Null Depth (ND) and Null Width (NW), which measure NW values at specified depths relative to the beampattern maxima (occurring at endfire direction). We investigate these measures as a function of depth for different DMA orders and beampatterns.

A null is defined as the local minima point of a beampattern, and in most cases for DMAs, it ideally corresponds to zero, leading to an ND value of $-\infty$ dB. However, in practice, actual ND values are significantly different and depend on various factors, such as the number of quantization bits, DMA order and pattern type. In this paper, we also examine the impact of signal quantization on linear DMAs, by formulating an analytical expression for the quantized beamformed output for any general $N^{th}$ order DMA. The paper also focuses on how signal quantization affects ND and NW in a beampattern. Results demonstrate that higher-order DMAs exhibit increased NW values, indicating a broadening of the null region with increasing order. Experiments were carried out to support the simulation-based findings. The resulting measured beampattern exhibits a significant ND value, reinforcing the validity and effectiveness in a practical experiment setup.

The remainder of the paper is organised as follows. Section II presents the signal model, quantisation analysis of the recorded signal for any general $N^{th}$ order DMA, followed by detailed analysis for the first-order dipole case, and finally, the introduction of null-related performance measures to evaluate the array performance. Section III presents simulation and experiment-based results along with the relevant discussion of the findings. The paper concludes with Section IV.

## II. SIGNAL MODEL, QUANTISATION ANALYSIS AND NULL-RELATED PERFORMANCE MEASURES

This section presents the signal model of our problem, followed by the analytical formulation of a general $N^{th}$ order DMA output for quantized signals. We also present the analytical derivation of the ND value for a $1^{st}$ order dipole case and compare it with the numerically obtained value. A detailed description of the null-related performance measures is provided in the subsequent subsection.

### A. Signal Model

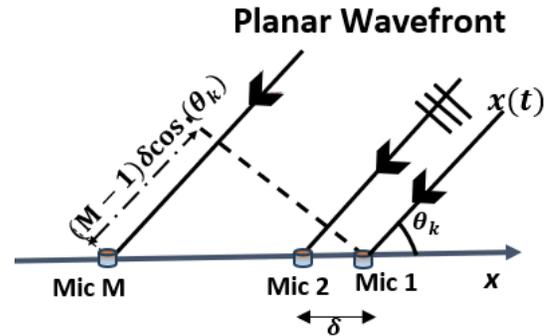

Fig. 1. Signal Model

A mathematical background of the problem is presented in this section. Let us consider a linear DMA of $M$ uniformly spaced omnidirectional mics incident by an acoustic source in a free-field environment, at an incident angle $\theta_k$ w.r.t. the mic array axis as shown in Fig.1. The signal of interest in the reference mic, i.e., mic-1 is given by [3]:

$$x(t) = A\cos(2\pi f_0 t + \phi_s) \quad (1)$$

where $A$, $f_0$ and $\phi_s$, are the amplitude, frequency, and initial phase of the incident sinusoidal signal respectively.

Let corresponding to $i^{th}$ mic, transfer function relating mic gain $G_i$ and phase $\phi_i$ at angular frequency $\omega_0 = 2\pi f_0$ is given by:

$$H_i(\omega_0) = G_i(\omega_0)\, e^{j\,\phi_i(\omega_0)} \quad (2)$$

where $i = 1,2, \ldots M$. The time domain signal recorded at the $i^{th}$ mic is thus given by:

$$y_i(t) = A\, G_i(\omega_0) \cos(2\pi f_0(t - \tau_i) + \phi_s + \phi_i(\omega_0)) \quad (3)$$

where $\tau_i = (i-1)\tau_0$, is the delay between the reference mic and the $i^{th}$ mic and $\tau_0 = \frac{\delta \cos(\theta_k)}{c}$, is the delay between two adjacent sensors at an incident angle $\theta_k$. Here, $\delta$ is the spacing between consecutive sensors and $c$ is the speed of sound.

The signal recorded is continuous in time and amplitude and is sampled and quantized into a digital signal using a data acquisition (DAQ) system.

### B. Signal Quantisation Analysis

For any $N^{th}$ order DMA, having $M$ ($=N+1$) sensors, the signal recorded at the $i^{th}$ mic is given by (3). The quantized and sampled version of this signal at sampling frequency $F_s$ is given as:

$$y_i^q[n] = Q(A\, G_i(\omega_0) \cos(2\pi f_0 (n - N_0)T_s + \phi_s + \phi_i(\omega_0))), \quad for\ i = 1, \dots M. \tag{4}$$

where $N_0 = (i-1)\frac{\tau_0}{T_s}$ is the discrete-time fractional delay corresponding to $\tau_i = (i-1)\tau_0$ and $T_s = \frac{1}{F_s}$ is the sampling time. Here $y_{i,in}^q[n] = y_i^q[n]$ is the in-phase component. The signal $Q(.)$ denotes the quantized representation of its argument.

The corresponding quadrature phase signal of $y_i^q[n]$ is:

$$y_{i,qp}^q[n] = Q(A\, G_i(\omega_0) \sin(2\pi f_0 (n - N_0)T_s + \phi_s + \phi_i(\omega_0))). \tag{5}$$

Let us represent the digital signal in complex form in terms of its in-phase and quadrature-phase components as:

$$z_i^q[n] = y_{i,in}^q[n] + j\, y_{i,qp}^q[n], \quad i = 1, \dots M. \tag{6}$$

To obtain the DMA beamformer output, complex weights $C_i$ are applied to each of the recorded signal channels where

$$\begin{aligned}C_i &= \left(\frac{1}{G_i(\omega_0)} e^{-j\phi_i(\omega_0)}\right) (D_i\, e^{j\psi_i}), \ i = 1, \dots M, \\ &= \frac{D_i}{G_i(\omega_0)} \{\cos(\psi_i - \phi_i(\omega_0)) + j\sin(\psi_i - \phi_i(\omega_0))\}. \end{aligned} \tag{7}$$

Here $D_i\, e^{j\psi_i}$ is the complex beamforming weight for the $i^{th}$ mic, where $D_i$ is the magnitude term and $\psi_i$ is corresponding phase term. $\frac{1}{G_i(\omega_0)} e^{-j\phi_i(\omega_0)}$ is the known inverse transfer function weight at the angular frequency $\omega_0$ to correct for the phase and gain mismatch between the mics. As would be the case with a DAQ, only the mic signal is quantized in integer/fixed point format here. Weight coefficients and all other arithmetic processing are in the native data type of the processor/software viz float/double format.

The complex signal output after weights multiplication for the $i^{th}$ mic channel is given as:

$$s_i^q[n] = C_i\, z_i^q[n]. \tag{8}$$

The expression for the quantized beamformed output is given as:

$$u^q[n] = \sum_{i=1}^M Re\,\{s_i^q[n]\} =$$

$$\sum_{i=1}^M \left\{ \left[\frac{D_i}{G_i(\omega_0)} \cos(\psi_i - \phi_i(\omega_0))\right] [Q(A.\,G_i(\omega_0) \cos(2\pi f_0 (n - N_0)T_s + \phi_s + \phi_i(\omega_0)))] - \left[\frac{D_i}{G_i(\omega_0)} \sin(\psi_i - \phi_i(\omega_0))\right] [Q(A.\,G_i(\omega_0) \sin(2\pi f_0 (n - N_0)T_s + \phi_s + \phi_i(\omega_0)))] \right\} \tag{9}$$

where $Re\{.\}$ denotes the real part of the argument.

We next estimate the beampattern which corresponds to the power of the quantized beamformed signal. The quantized signal can be expressed as a sum of the quantization error and the unquantized signal. Therefore, the beamformed output due to the quantization of the signal can be written as sum of (i) terms introduced due to the unquantized signal processing and (ii) terms introduced due to the processing of the error during signal quantization. The argument enclosed in [.] represents the unquantized part.

For any general $N^{th}$ order DMA, (9) can therefore be rewritten for the quantized beamformed output as:

$$\begin{aligned}u^q[n] &= \sum_{i=1}^M \Big\{ \left[\frac{D_i}{G_i(\omega_0)} \cos(\psi_i - \phi_i(\omega_0))\right].\left[A\,G_i(\omega_0) \cos(2\pi f_0 (n - N_0)T_s + \phi_s + \phi_i(\omega_0)) + \varepsilon_{i,in}[n]\right] \\ &\quad - \left[\frac{D_i}{G_i(\omega_0)} \sin(\psi_i - \phi_i(\omega_0))\right].\left[A\,G_i(\omega_0) \sin(2\pi f_0 (n - N_0)T_s + \phi_s + \phi_i(\omega_0)) + \varepsilon_{i,qp}[n]\right] \Big\} \\ &= \sum_{i=1}^M \left\{ \begin{matrix} A\,D_i \cos(2\pi f_0 (n - N_0)T_s + \phi_s + \psi_i) + \\ \frac{D_i}{G_i(\omega_0)} \cos(\psi_i - \phi_i(\omega_0))\, \varepsilon_{i,in}[n] - \\ \frac{D_i}{G_i(\omega_0)} \sin(\psi_i - \phi_i(\omega_0))\, \varepsilon_{i,qp}[n] \end{matrix} \right\} \end{aligned} \tag{10}$$

where $\varepsilon_{i,in}[n]$ is the quantization error in the $i^{th}$ mic channel for in-phase component (i.e. as recorded) and $\varepsilon_{i,qp}[n]$ is the quantization error in the $i^{th}$ mic channel for the quadrature phase component. Both error components are assumed to be independent and can be reasonably approximated by the uniform density functions $U\left(-\frac{\Delta_{in}}{2}, \frac{\Delta_{in}}{2}\right]$ and $U\left(-\frac{\Delta_{qp}}{2}, \frac{\Delta_{qp}}{2}\right]$ respectively. Here $\Delta_{in}$ is the quantization step-size for the in-phase signal component while $\Delta_{qp}$ is the quantization step-size for the quadrature-phase signal component. The in-phase component can be used to generate its quadrature counterpart either in the analog domain, before digitization, using analog electronics, or derived in the digital domain via the Hilbert transform.

The expression for the beampattern $B$ is then given as:

$$B(\theta_k) = \frac{1}{L}\sum_n |u^q[n]|^2 \tag{11}$$

where $n = 1, 2, \ldots\ldots..L$ where $L$ is the length of sequence.

**Analysis for a 1st order Dipole**

For a 1st order dipole i.e., $M = 2$, we have $\psi_1 = \frac{\pi}{2}$ and $\psi_2 = -\frac{\pi}{2}$ respectively. Without loss of generality, microphone array gains $G_i(\omega_0)$ and sinusoidal amplitude $A$ are taken as unity.

Further for the 1st order dipole case, $D_1 = D_2 = D = 3.98$ [3]. Using (10), the quantized signal's beamformed output expression for a 1st order dipole can be reduces to:

$$u^q[n] = D \cos\left(2\pi f_0 n T_s + \phi_s + \frac{\pi}{2}\right) + \\ D \cos\left(2\pi f_0 (n - N_0) T_s + \phi_s - \frac{\pi}{2}\right) + D \cos\left(\frac{\pi}{2} - \phi_1\right)\varepsilon_{1,in}[n] - D \sin\left(\frac{\pi}{2} - \phi_1\right)\varepsilon_{1,qp}[n] + D \cos\left(\frac{\pi}{2} - \phi_2\right)\varepsilon_{2,in}[n] - D \sin\left(\frac{\pi}{2} - \phi_2\right)\varepsilon_{2,qp}[n] \quad (12)$$

where $N_0 = (i - 1)\frac{\delta \cos(\theta_k)}{c\, T_s}$ is the discrete-time fractional delay corresponding to $\tau_i$. Now, the overall power of this quantized beamformer output $u^q[n]$ is subsequently computed by varying the source incident angle $\theta_k$ over the entire azimuthal range from 0 to $2\pi$. Thus, the expression for the beampattern for a dipole is given as:

$$B_{dip}(\theta_k) = E\{|u^q[n]|^2\}. \quad (13)$$

The beampattern plots for all distinct orders and patterns are generated by performing normalization with respect to the maxima, which is set to 0 dB.

At null location i.e. when $\theta_k = \theta_{Null} = \frac{\pi}{2}$, the sum of the first two terms in the above expression (12), is equal to zero, which corresponds to the ideal case of no signal quantization, that is ideally shown in beampattern plots. However, the other terms contribute to the error introduced because of the quantization of the signal. Thus expression for the quantized signal's beamformed output for a 1st order dipole at null location reduces to:

$$u^q[n]\ _{at\ \theta_{Null}} = D \cos\left(\frac{\pi}{2} - \phi_1\right)\varepsilon_{1,in}[n] - \\ D \sin\left(\frac{\pi}{2} - \phi_1\right)\varepsilon_{1,qp}[n] + D \cos\left(\frac{\pi}{2} - \phi_2\right)\varepsilon_{2,in}[n] - D \sin\left(\frac{\pi}{2} - \phi_2\right)\varepsilon_{2,qp}[n]. \quad (14)$$

Assuming that the in-phase and corresponding quadrature-phase error terms are independent, the cross-product terms in the power expansion of (14) become negligible and can be treated as zero. Also, since (14) involves a linear combination of four independent error terms, each following a uniform distribution, therefore, the overall distribution involves the convolution of these 4 uniform density functions with corresponding mean equal to the weighted sum of the individual means, and a variance equal to the sum of the scaled individual variances, based on the coefficients in the linear combination. Thus, the expression for ND for a 1st order dipole pattern is:

$$ND_{dipole} = B_{dip}(\theta_{Null}) = E\left\{\left|u^q[n]\ _{at\ \theta_{Null}}\right|^2\right\} \\ = D^2 \cos^2\left(\frac{\pi}{2} - \phi_1\right)\varepsilon_{1,in}^2[n] + \\ D^2 \sin^2\left(\frac{\pi}{2} - \phi_1\right)\varepsilon_{1,qp}^2[n] + D^2 \cos^2\left(\frac{\pi}{2} - \phi_2\right)\varepsilon_{2,in}^2[n] + D^2 \sin^2\left(\frac{\pi}{2} - \phi_2\right)\varepsilon_{2,qp}^2[n]. \quad (15)$$

Now, solving the above expression and normalized for a 16-bit quantization depth level, ND for the 1st order dipole case is coming out to be -83.1 dB.

### C. Null Related Measures

The commonly used measures in literature to study the performance of any differential beamformer are the DF and WNG [18]-[26].

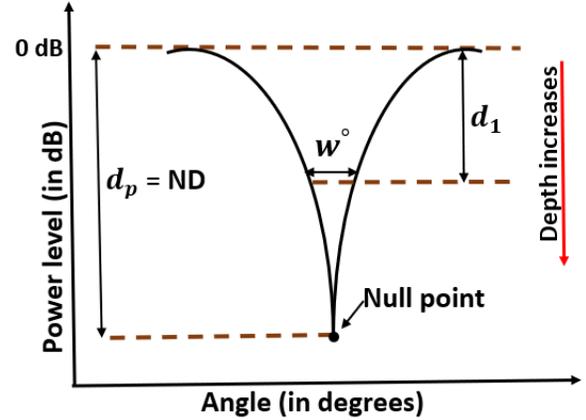

Fig. 2. Beampattern plot to represent ND and NW(*d*)

The DF measures the ratio of a beamformer's response in the end-fire direction ($\theta = 0°$) to its average response across all directions, while WNG assesses the beamformer's robustness against imperfections like sensor noise and positional errors.

These conventional DMA performance measures, such as DF and WNG, primarily focus on the main lobe characteristics and do not effectively capture the behaviour of beampattern nulls. To address this, we introduce null-specific measures viz ND and NW as function of depth (NW*(d)*), which quantify the depth and angular spread of nulls relative to the maxima of the beampattern i.e. 0 dB as shown in Fig.2. These measures are analysed across different DMA orders and pattern types. The subsequent section briefly presents the values of each parameter.

1. **Null Depth (ND)** is a quantitative measure that indicates how effectively a directional array, such as DMA, can suppress signals arriving from specific directions. In the context of a beampattern, a *null* refers to a local minimum in the spatial response, typically located in a direction where signal rejection is desired. Ideally, this minimum corresponds to zero gain, leading to an ND value of negative infinity ($-\infty$ dB). However, in practical scenarios, several factors such as hardware limitations, the number of signal quantization bits, and array imperfections prevent the null from reaching this theoretical limit. As a result, the actual ND observed is finite and varies depending on the beam pattern, order, and number of signal quantization bits. A deeper null (larger negative value in dB) indicates stronger suppression capability and is particularly important in scenarios involving stronger interfering sources. ND corresponds to the depth value at the null point i.e. $d=d_p$ as shown in Fig. 2.

2. **Null Width (NW)**, on the other hand, refers to the degree of spread between two adjacent lobes in a beampattern and measures the width of the null at specified depth levels relative to the maxima. It provides insight into the sharpness or spatial selectivity of a null. A small NW indicates that the array has a high angular resolution around the null direction, which is desirable for precisely targeting interference or noise sources. Conversely, a large NW suggests a more diffuse null, which may reduce its effectiveness in applications requiring precise spatial filtering. NW is influenced by the same factors as ND, such as the DMA order, quantization bits, and beampattern shape and typically exhibits an inverse relationship with ND. That is, deeper nulls often correspond to narrower NW values, especially in lower-order arrays, which are capable of forming sharper and more defined nulls. In Fig.2, we see NW($d$)= $w°$ at depth $d=d_1$. When NW($d_p$) > 0, the null is said to be broadened. We investigated ND and NW($d$) measures in detail for various DMA orders and beampatterns.

### III. SIMULATION AND EXPERIMENT-BASED STUDIES AND DISCUSSION

In this section, we present both simulation and experimental results to illustrate ND and NW($d$) for various beampatterns across different DMA orders, along with their variation as a function of signal quantization bit depth.

#### A. Simulation setup and studies

In this section, we study the impact of DMA signal quantization over narrowband tonal signals. For simulations, the source frequency is taken as 997 Hz (prime number, irrationally related to sampling frequency). This is done to avoid spectral leakage artifacts or unintentional periodicity in simulations and analysis. The sampling frequency considered is 44.1 kHz, inter-element spacing between adjacent sensors is kept as 13.76mm ($\delta = 0.04\lambda$). A total of 5000 Monte Carlo runs are used to calculate the average ND and NW($d$) values for different orders and beampatterns of dipole, cardioid, hyper-cardioid, and super-cardioid. In each run, $\phi_i$ and $\phi_s$ are randomly generated between 0 to $2\pi$. MATLAB simulation results are presented to support the discussion.

ND values for 1st, 2nd and 3rd order DMA for dipole, cardioid, hyper-cardioid, and super-cardioid beampatterns are presented in Table I. For example, with a 16-bit quantization depth, a first-order dipole achieves an ND of -83.1 dB (w.r.t end-fire). This simulated result matches the analytically derived ND for a first-order dipole (as shown in the above section), thereby validating the correctness of the analytical modelling. MATLAB simulation results show that increasing the number of quantization bits results in deeper nulls (larger negative value in dB), indicating that interfering signals coming from null directions can be suppressed more efficiently. Fig. 3 depicts the ND plot for different DMA orders and beampatterns. For every DMA order, the ND values progressively deepen with higher quantization bits, gradually converging toward the values observed in the unquantized scenario.

Since 96 kHz is a commonly used sampling rate for high-resolution audio due to its increased precision and ability to capture finer waveform details, the simulation was also carried out at 96 kHz. Increasing the sampling rate from 44.1 kHz to 96 kHz showed similar behaviour in terms of ND and NW($d$). However, a variation in ND was observed of approximately 1 dB for lower quantization levels (16 bits) and up to 5 dB for higher quantization levels (20 and 24 bits). This difference can be attributed to the more accurate representation of the waveform at higher sampling rates. Moreover, nonlinearities in quantization and imperfect reconstruction filters can influence the accuracy of null formation, particularly at higher bit depths where the system becomes more sensitive to subtle artifacts.

TABLE I
Null Depth (in dB) as a function of Beampattern type and Quantization Bits

| DMA order and beampatterns | ND (in dB) | | | | |
|---|---|---|---|---|---|
| | Unquan-tized | No. of quantization bits | | | |
| | | 12-bit | 16-bit | 20-bit | 24-bit |
| **1st order (M=2)** | | | | | |
| Dipole (Null at 90°) | -239.1* | -58.9 | -83.1 | -107.1 | -131.2 |
| Cardioid (Null at 180°) | -245.7* | -64.8 | -88.5 | -113.1 | -137.2 |
| Hypercardioid (Null at 120°) | -236.8* | -62.4 | -85.1 | -110.6 | -134.7 |
| Supercardioid (Null at 135°) | -242.8* | -63.5 | -86.2 | -111.7 | -135.8 |
| **2nd order (M=3)** | | | | | |
| Dipole (Null at 90°) | -223.4* | -42.2 | -66.3 | -90.4 | -114.5 |
| Cardioid (nulls at 90°, 180°) | -229.2*, -222.4* | -48.1, -48.1 | -72.3, -72.3 | -96.3, -96.3 | -120.4, -120.4 |
| Hypercardioid (nulls at 72°, 144°) | -96.3, -96.3 | -44.1, -44.1 | -68.2, -68.2 | -90.8, -90.8 | -96.2, -96.2 |
| Supercardioid (nulls at 106°, 153°) | -115.2, -115.5 | -49.7, -49.7 | -73.6, -73.8 | -97.6, -97.8 | -115.0, -115.3 |
| **3rd order (M=4)** | | | | | |
| Dipole (Null at 90°) | -203.1* | -24.9 | -49.0 | -73.1 | -97.2 |
| Cardioid (null at 90°, 180°) | -213.8 *, -206.5* | -25.1, -25.2 | -52.4, -52.4 | -73.3, -73.3 | -97.4, -97.4 |
| Hypercardioid (nulls at 55°,100°,145°) | -76.5, -76.4, -76.2 | -24.1, -24.1, -24.1 | -48.2, -48.2, -48.1 | -70.9, -70.8, -70.6 | -76.4, -76.3, -76.1 |
| Supercardioid (nulls at 97°,122°,153°) | -87.8, -87.4, -87.4 | -35.1, -35.1, -35.1 | -58.2, -58.2, -58.1 | -83.3, -83.4, -83.2 | -87.7, -87.3, -87.3 |

* Computed using double precision arithmetic in MATLAB, having a range of the order of $10^{-308}$, which approximates to ideal negative infinite in dB.

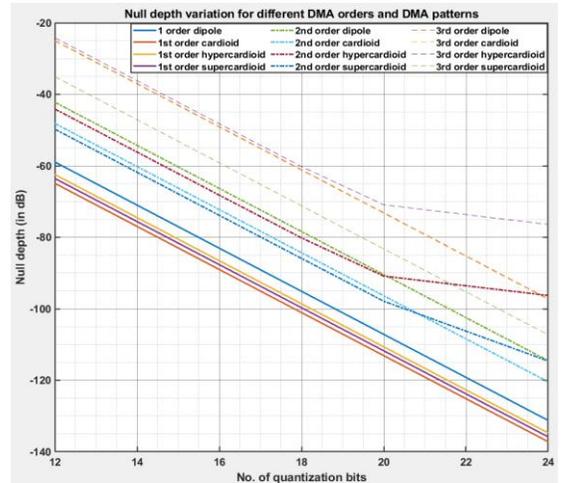

Fig. 3. Null depth as a function of quantization bits for different DMA orders and patterns.

TABLE II
NULL WIDTH AS A FUNCTION OF DEPTH FOR DIFFERENT DMA ORDERS AND PATTERNS

| Depth ($d$) in dB | NW ($d$) in degrees | | | | | |
|---|---|---|---|---|---|---|
| | 1st order | | 2nd order | | 3rd order | |
| | Unquantized | 16-bit quantized | Unquantized | 16-bit quantized | Unquantized | 16-bit quantized |
| | Beampatterns: [dipole ; cardioid ; hypercardioid ; supercardioid] | | | | | |
| -10 | [37.7°; 136.1°; 72.7°; N.A.] | [36.0°; 136.0°; 69.4°; 79.1°] | [68.6°; N.A. ; N.A. ; N.A.] | [68.0°; N.A. ; N.A. ; N.A.] | [86.1°; N.A. ; N.A. ; N.A.] | [86.0°; N.A. ; N.A. ; N.A.] |
| -20 | [11.8°; 73.3°; 21.9° ; 28.7°] | [11.5°; 73.0°; 19.0° ; 28.0°] | [36.4°;26.0°;88.9°; 13.6°,23.9°; N.A.] | [36.2°;25.8°;88.0°; 13.0°,23.0°; N.A.] | [56.0°; N.A. ; 23.0°, 24.2°, 12.9° ;N.A.] | [55.8°; N.A. ;22.7°, 24.0° ,12.0°; N.A.] |
| -30 | [3.9° ; 40.7° ; 6.3° ; 8.6°] | [3.6° ; 40.0°; 6.0° ; 8.0°] | [18.1°;7.8°,44.0°; 4.3°,7.8°;16.4°,39.1°] | [18.0°;7.1°,40.9°; 4.2°,6.9°;16.0°,39.0°] | [36.5°; 30.2°,44.2°; 6.6°,7.2°,3.8°; N.A. ] | [36.2°;30.0°,44.0°;6.0° , 7.0°,3.6°; N.A.] |
| -40 | [1.1° ; 22.8° ; 1.9° ; 2.7°] | [0.9° ; 22.0°; 1.8° ; 2.5°] | [8.0°;2.8°,24.0°; 1.3°,2.1°;4.7°,10.0] | [7.9°;2.6°,22.0°; 1.4°,2.4°;4.6°,9.8°] | [24.4°; 2.2°, 24.0°; 2.0°, 2.3°, 2.1°; N.A.] | [24.1°; 16.9°,23.0°; 1.9°,1.9°,2.0°;N.A.] |
| -50 | [0.4° ; 12.8°; 1.5° ; 2.0°] | [0.2° ; 12.0°; 1.0° ; 1.9°] | [2.8°; 2.0°,13.4°; 0.4°,1.5° ;3.0°, 4.0°] | [2.0°;1.6°,12.8°;0.4°,0.6°;1.4°,3.0°] | [17.0°; 9.0°, 13.0°;1.4°, 1.4°,1.0°; 9.3°,9.5°, 4.0°] | [N.A. ; 8.3°,11.6°; N.A., N.A.,N.A; 8.5°,8.7°,3.5°] |
| -60 | [0.1° ; 7.2° ; 1.2° ; 1.3°] | [0.1° ; 6.9°; 0.2° ; 1.0°] | [1.8°;1.0°,8.0°; 0.2°,1.3°;0.5,2.0°] | [1.6°;0.6°,7.0°;0.1°, 0.5°; 0.5, 1.0°] | [11.7°; 6.0°,7.2°; 1.0°, 0.8°,0.2°; 3.0°,3.0°,1.6°] | [N.A.; N.A.; N.A. ; N.A.] |

Figures 4-6 show beampattern plots for 1st, 2nd and 3rd order DMA, respectively. It is evident from these plots that ND values are not negative infinity but are limited based on factors such as the quantization effect, DMA order, and type of beampatterns. This insight is not evident from the ideal plots that we see in the literature.

Table. II above shows NW values for different depths $d$ varying between -10dB to -60dB at an interval of 10dB each for different DMA orders and patterns of dipole, cardioid, hypercardioid, and supercardioid patterns. ND values are calculated w.r.t the beampattern's maxima (i.e., 0 dB). The table compares the unquantized (ideal) case values with a 16-bit quantized value. Every $N^{th}$ order DMA has a maximum of $N$ nulls in its beampattern plot. For each $d$, NW($d$) for all $N$ nulls (separated by commas) and for each different beampattern (distinguished by semicolons) is presented in the table. It is observed from the table that lower-order DMAs exhibit smaller NW values compared to higher orders. The "N.A." entry in the table denotes "Not Applicable," indicating the absence of two adjacent lobes in the beampattern plot, and consequently, the NW cannot be defined at that depth level.

This suggests that deeper and more pronounced nulls are typically associated with narrower widths, particularly in lower-order arrays, which are inherently better at producing sharper and more localized nulls. For applications where broader nulls are required, higher order DMA can be useful to mitigate interference coming from a wider range of angles (e.g., noise from the window of a car, recorded at the dashboard for a car infotainment system), as they offer null broadening. Thus, the trade-off between ND and NW values emerges as an important performance factor depending on the specific application requirements.

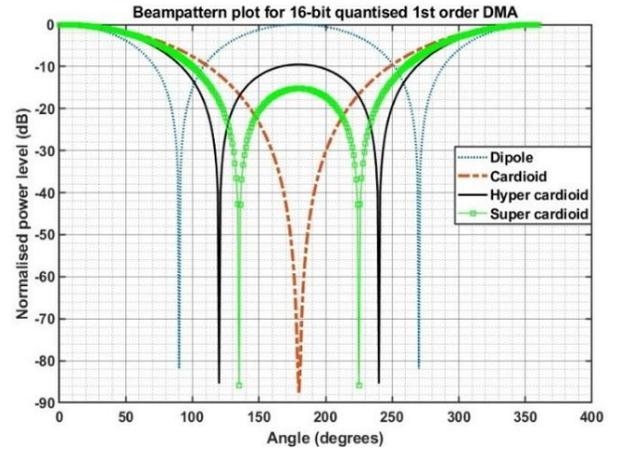

Fig. 4. Beampattern for 1st order quantized DMA

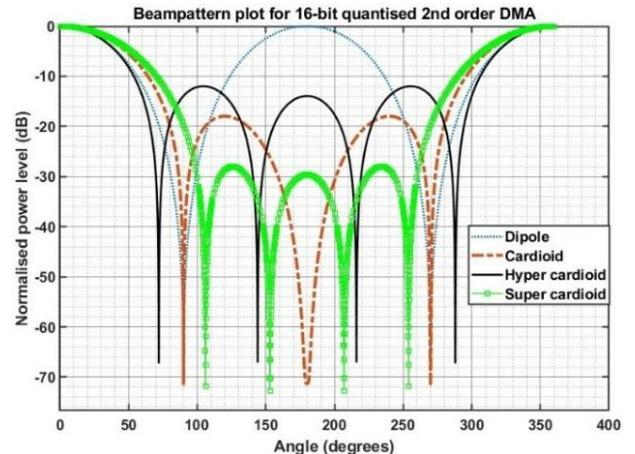

Fig. 5. Beampattern plot for 2nd order quantized DMA

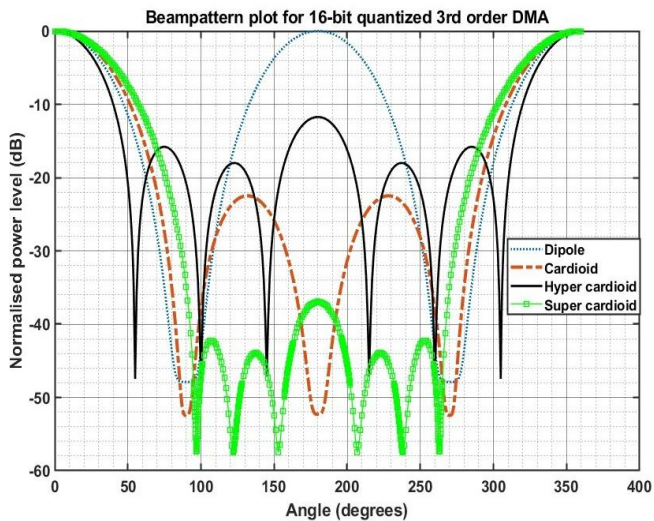

Fig. 6. Beampattern plot for 3$^{rd}$ order quantized DMA

### B. Experimental Studies

This subsection presents experiments using recorded signals to investigate signal quantization on linear differential beamformers. The experiments were conducted in a fully anechoic room of size 3.34m x 3.04m x 2.06m in our laboratory. We initially built a microphone array holding structure (as shown in Fig.7 and Fig.8) using Polymethyl Methacrylate (PMMA) plastic. This material can withstand a boiling point of up to 200°C, making it suitable for use in applications such as automobile dashboards, where temperatures are relatively elevated. Experiments are performed to present measured beampattern results for 1$^{st}$ and 2$^{nd}$ order linear DMA structures constructed using MKE 1 Sennheiser omnidirectional microphones. A Beheringer U-PHORIA Audiophile 8-channel Mic preamplifier with an integrated ADC is used to process the microphone outputs. The outputs of the microphones are transformed into digital signals at 44.1 kHz sampling rate. A loudspeaker (JBL NANO K3 series) positioned 1.1m in front of the DMA array acts as the source of the desired frequency. The experimental setup inside the anechoic room is shown in Fig. 9. The optimum interelement spacing between two adjacent microphones has been taken slightly larger than the spacing used in the simulation ($\delta_{opt}$ = 15.64 mm), due to the physical size constraints of the microphones. The calibration between microphones has been done as explained above.

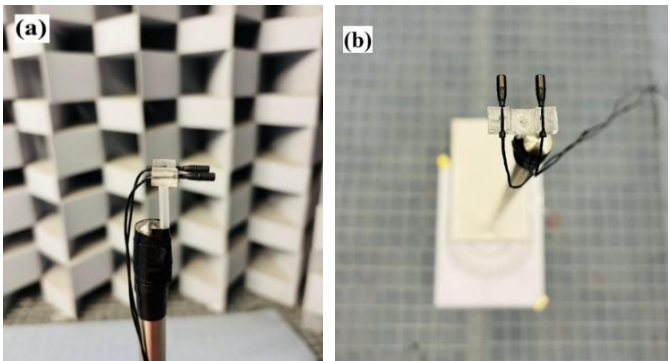

Fig. 7. Two-Microphone array holding structure:
(a) Side view (b) Top view

To obtain the results for the measured beampattern, the source loudspeaker is made to rotate at angles from 0° to 360°, incremented in a stepsize of 10° each. Angles are made to vary by 1° near the null regions to obtain a more precise pattern response. Both the array and the loudspeaker are on the same horizontal plane, which is 79.5 cm above the sound absorption floor, as shown in Fig. 9. The center of the array is coincident with the center of the loudspeaker. The loudspeaker plays the sinusoidal signal of the specified frequency. To obtain the measured beampattern, the gain is computed in every rotated direction based on the beamformer's output. In the following section, we present measured results.

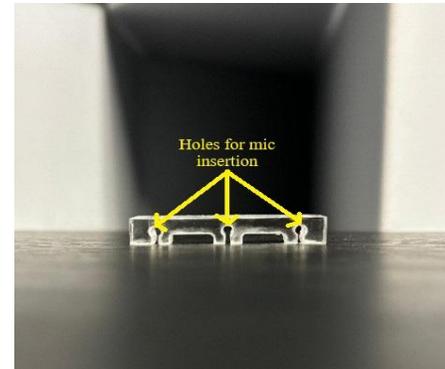

Fig. 8. Three-microphone array holding structure (side view)

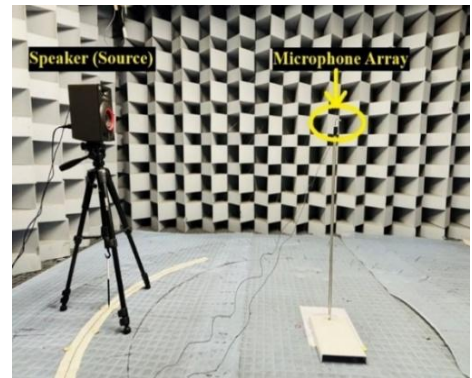

Fig. 9. Experimental setup for measuring beampattern in fully anechoic room

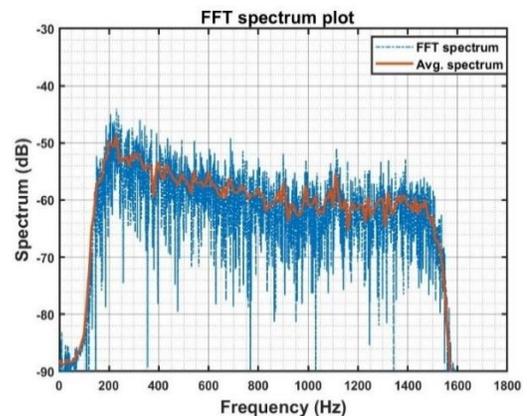

Fig. 10. Spectrum of differential noise recorded between two channels for a 1$^{st}$ order cardioid pattern

We conducted two sets of experiments to show the measured beampattern for the narrowband source case. In the first experiment, recordings are done to measure beampattern for a 1st order cardioid pattern using the two microphone array holding structure (shown in Fig.7). Silence within the anechoic room was also recorded prior to the beamforming signal being recorded, providing a noise floor that serves as a reference level for the reported ND value.

Fig.10 presents a spectrum plot of the differential noise recorded between two microphone channels in the presence of both environmental and electronic noise (e.g., when the loudspeaker is powered on). Each recorded signal was processed using an FIR bandpass filter to suppress harmonic frequencies as well as other extraneous frequency components present in the signal spectrum. Fig.11a. and Fig.11b. shows the spectrum plot of beamformed signal recorded at 180°(null location) and 0° (endfire) respectively. From the figures, it is inferred that the 997 Hz signal from the 180° direction has been completely attenuated, thereby showing the efficacy of the differential beamformer characteristics. Fig.12 compares the simulated and measured beampatterns at the 997 Hz frequency bin for a 16-bit quantization depth. It shows that for a 1st order cardioid pattern, the measured ND value reaches a depth of -68.7 dB (The depth values are determined w.r.t. power along endfire, i.e. 0° direction). Although this value does not match the simulation's expected -88.5 dB depth, it demonstrates a high degree of agreement and validates the directional suppression capability of the beamformer.

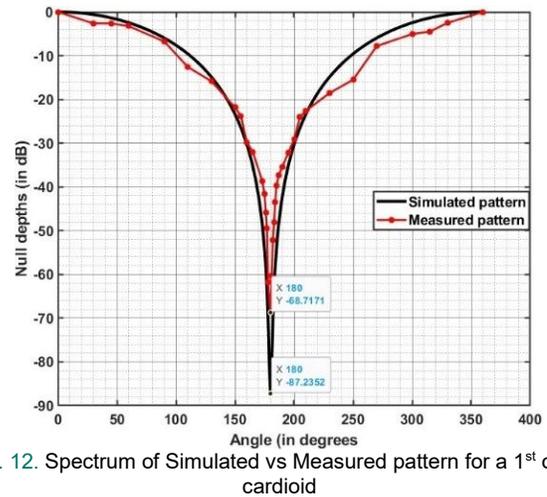

Fig. 12. Spectrum of Simulated vs Measured pattern for a 1st order cardioid

The second experiment measures the beampattern for a 2nd order cardioid pattern with nulls at 90° and 180°, respectively, using the three microphone array holding structure (shown in Fig.8). Spectrum plots for differential noise recorded at array output and recorded beamformed signal are displayed in Fig. 13 and Fig. 14, respectively. ND values are calculated relative to the endfire direction (0°). Experimental results show that null depths of up to -35.5 dB at 90° and -35.6 dB at 180° are achieved. Fig.15 plots the designed and measured beampatterns corresponding to a 2nd order cardioid pattern at 997Hz bin for a 16-bit quantization depth. From the figures, it has been inferred that even though the ND for the measured pattern is not as sharp as that of the simulation pattern, the measured beampattern still indicates significant ND.

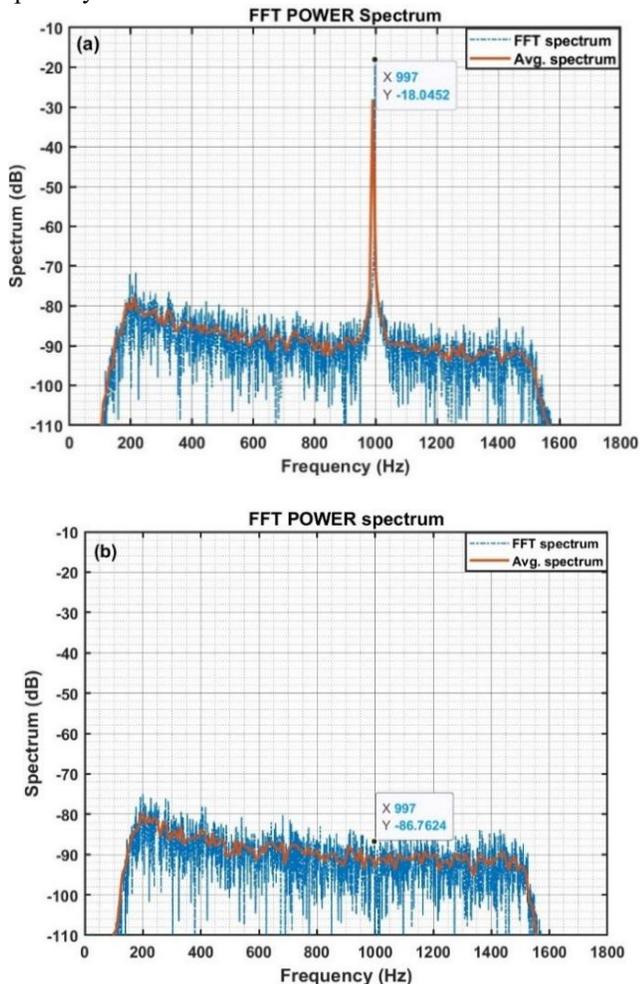

Fig. 11. Spectrum of: (a) beamformed signal recorded at 0° (b) beamformed signal recorded at 180°

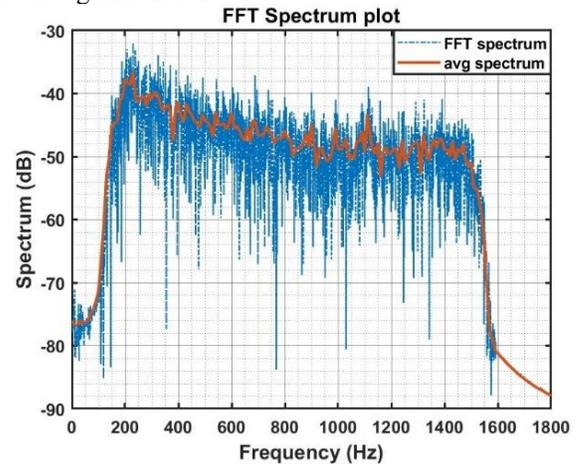

Fig. 13. Spectrum of differential noise recorded between three channels for a 2nd order cardioid pattern

In general, the measured beampattern is close to the designed (simulated) beampattern, however, the nulls of the measured pattern are not as deep as those in the designed beampattern. This discrepancy between simulation and experimental results can be attributed to several factors. First, we used a loudspeaker as a source. The loudspeaker measures 12.2 cm in width and 20.9 cm in height, with a 7.1 cm inner diaphragm. Because the loudspeaker has a finite size, it emits sound from an extended area rather than a single point, thus deviating from the idealized point source assumed in the theoretical model. Second, the ND value obtained in the measured patterns is constrained by the noise floor associated with each order. Finally, the angular resolution of the measurements was limited to 10-degree increments. As observed from the simulated beampattern, even a 1-degree deviation in measurement angle can introduce an error of approximately 7 dB, thereby affecting the accuracy of the observed ND values in real recordings. Despite these limitations, the measured beampattern still demonstrates a significant ND, validating the effectiveness of the experimental setup.

featuring a null at 180°, and the second-order pattern exhibiting two nulls each at 90° and 180°. For each case, the ND corresponding to 997Hz is given, assuming spectral leakage in the adjacent bin is negligible. The depth values are determined w.r.t. power along endfire, i.e. $0°$ direction ($P0°$). Additionally, the table also includes NW values at various depth levels relative to 0 dB for both first- and second-order measured beampatterns. The calculated power values are normalized with respect to dB overload.

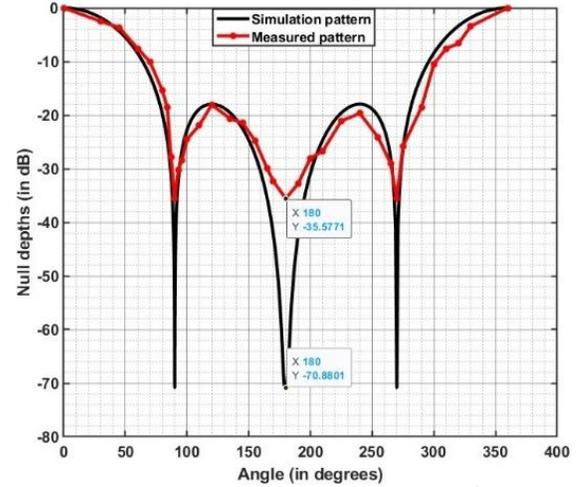

Fig. 15. Simulated vs Measured beampattern for a 2$^{nd}$ order cardioid pattern

TABLE III
Experimental results for Null Depth and Null Width variations for 1$^{st}$ and 2$^{nd}$ order cardioid beampattern

| S. No. | Parameters | 1$^{st}$ order Null at 180° | 2$^{nd}$ order Null at 90°; 180° |
|---|---|---|---|
| 1. | Noise floor level (in dB) | -68.9 | -56.9 |
| 2. | Power level at 997Hz (in dB) | | |
| | 0°  ($P0°$) | -18.1 | -35.0 |
| | 90°  ($P90°$) | -- | **-70.5** |
| | 180°  ($P180°$) | -86.8 | -70.6 |
| 3. | Corresponding ND (in dB) at | | |
| | 90°  ($P90° - P0°$) | -- | **-35.5** |
| | 180°  ($P180° - P0°$) | -68.7 | -35.6 |
| 4. | NW w.r.t 0dB (in degrees) | | |
| | Depth level : -10dB | 162.9° | -- |
| | -15dB | 126.3° | -- |
| | -20dB | 78.8° | **31.2°** ;104.1° |
| | -25dB | 39.1° | **11.5°** ; 57° |
| | -30dB | 38° | **5.5°** ;26.7° |

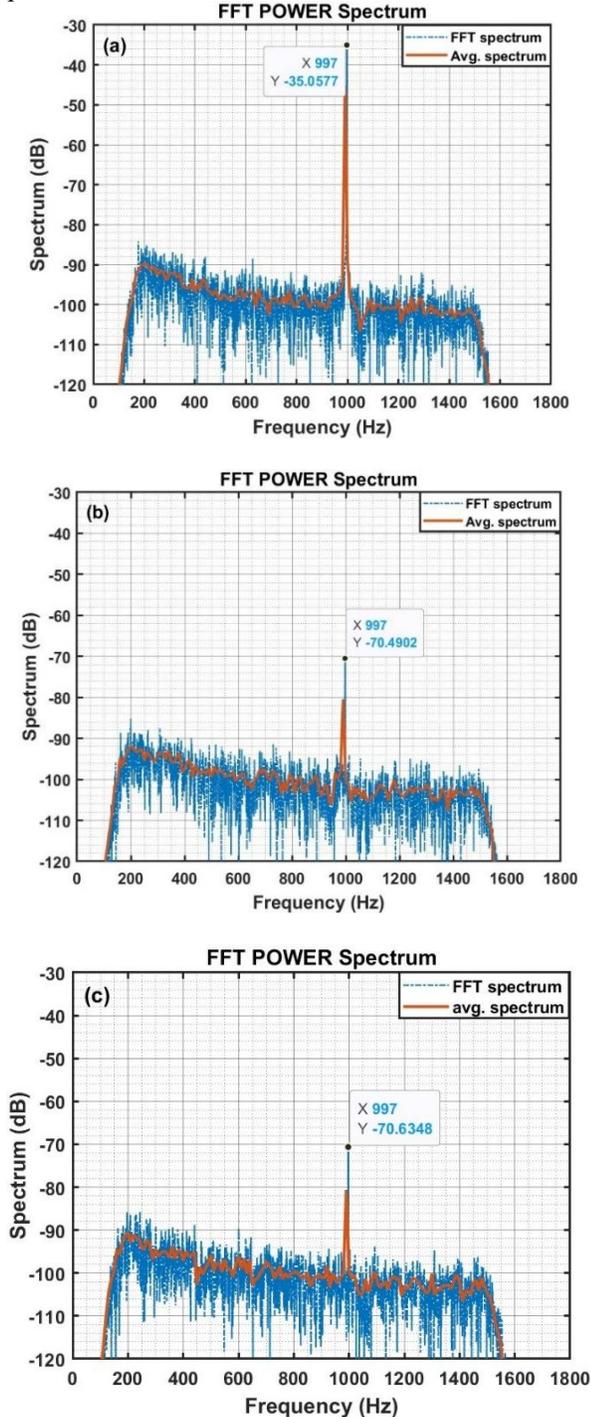

Fig. 14. Spectrum of: (a) beamformed signal recorded at 0° (b) beamformed signal recorded at 90° (c) beamformed signal recorded at 180°

Table.III summarises the experimental results for ND values for 1$^{st}$ and 2$^{nd}$ order cardioid patterns, with the first-order pattern

## IV. CONCLUSION

Existing DMA metrics like DF and WNG primarily capture peak-related characteristics of the beampattern, offering limited insight into the effectiveness of nulls. This paper re-examines the relevance of the DMA applications from the perspective of nulling of interfering sources. The study evaluates the performance of DMAs using specific null-based metrics, which are critical for characterizing the array's capability to suppress interfering sources. A null is a local minimum in the beampattern, ideally giving ND values of $-\infty$ dB, but in practice, ND values vary significantly with parameters such as quantization bits, DMA order, and beampattern type. The work further investigates the influence of microphone signal quantization on these null properties by examining variations across different DMA orders and beampattern types. The analysis focuses on null-related measures, relevant for applications aimed at nullifying interfering sources, and these are systematically examined in detail for 1$^{st}$, 2$^{nd}$ and 3$^{rd}$ order DMAs, as well as for standard beampatterns including dipole, cardioid, hypercardioid, and supercardioid configurations. An analytical formulation to study the impact of microphone signal quantization on the beamformer output for any general $N^{th}$ order DMA is presented. The dependence of ND on the number of quantization bits is also studied. The results indicate that higher DMA orders have larger NW($d$) values thus resulting in null broadening for increasing DMA orders. Lab experiments are performed in a fully anechoic room that corroborates the simulation studies. The validity and efficacy of the experimental setup are further supported by the obtained measured beampatterns, which show a significant ND value.